# Direct Dark Matter Search using CCDs

J. Estrada[1] for the DAMIC Collaboration
Fermilab Center for Particle Astrophysics
MS 127, POBox 500, Batavia, IL, 60510, USA

E-mail: estrada@fnal.gov

**Abstract.** There is currently vast evidence for Dark Matter (DM) from astronomical observations. However, in spite of tremendous efforts by large experimental groups, there is no confirmed direct detection of the dark matter in our galaxy. Recent experimental results and theoretical developments suggest the possibility of a DM particle with mass below 10 GeV, such a particle would escape most of the direct searches due to the large thresholds for the detection of nuclear recoils typically used. In this work we study the possibility of a new Dark Matter search with an unprecedented low threshold for the detection of nuclear recoils using high-resistivity CCD detectors (hr-CCD). Due to their extremely low readout noise and the relatively large active mass, these detectors present a unique opportunity in this field.

The Weakly Interactive Massive Particle (WIMP) is the leading candidate dark matter particle. The search for dark matter in the halo of our galaxy by direct detection experiments has been a very active field in recent years, with the limits on WIMP-nucleon cross section improving significantly with time (for a review see Refs. [1] and [2]). However, typical searches of this kind have a poor sensitivity for low mass dark matter particles because of the high thresholds (~1 keV) established for the detection of the nuclear recoils. The experimental limits obtained for dark matter searches rise sharply as the particle mass approaches 10 GeV, as discussed below (see Fig. 5). One of the limitations for setting a lower threshold in the direct dark matter searches is the readout noise of the detectors used. There has been a continuing effort in reducing the readout noise for the detectors typically used in direct dark matter searches to overcome the high threshold issue [3].

The most popular models for dark matter particles predict masses above 50 GeV (for a review see Ref. [4]), a mass high enough that low energy thresholds would not be needed. However, there are models [5][6] where the dark matter particles have low mass and for which their detection will require a low energy threshold for nuclear recoils. A low mass dark matter candidate could also arise in the most simple extension of the Minimal Supersymmetric Standard Model as discussed in Ref. [7] and Ref. [8]. Models for which the typical velocity of the dark matter particles with respect to the earth is lower have been considered [9][10], and those cases also require a low threshold in a direct search experiment. In addition to the theoretical motivations for a low mass dark matter search, the recent results from the DAMA/Libra collaboration could also be interpreted as the evidence for a low mass dark matter particle. In particular, it has been pointed out that the DAMA/Libra annual modulation signal is consistent with a low mass dark matter particle if channeling is occurring in the detector. This channeling would also make all other null results reported for DM direct searches compatible with the DAMA/Libra observations [11].

We are developing a low threshold and low background experiment with CCD detectors for a dark matter search, such detectors will allow a search optimized for low mass particles. The CCD detectors the we propose to use are the high resistivity fully depleted CCDs developed for astronomy due to their higher efficiency in the near-IR [12]. Fermilab, is leading the construction of the DECam instrument for the Dark Energy Survey [13], this is currently the largest instrument being built using these type of detectors. In the process of building DECam we have acquired comprehensive expertise on the high resistivity CCDs, their readout electronics and their packaging.

---

[1] To whom any correspondence should be addressed.

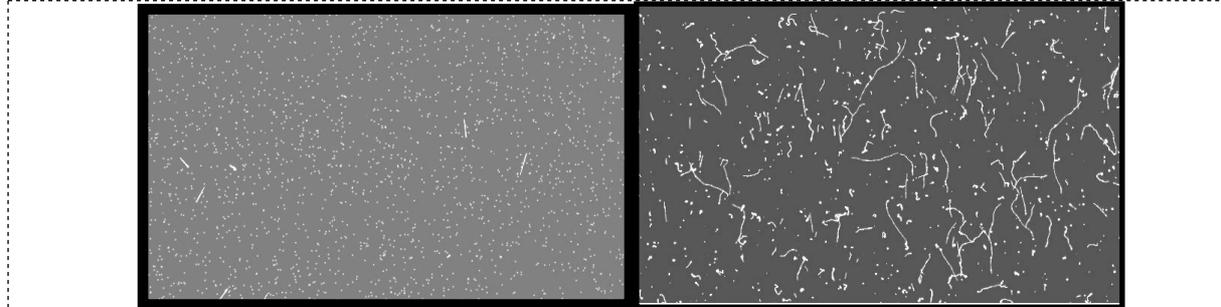

**Figure 2.** CCD exposure to $^{55}$Fe X-Ray source (left), and $^{252}$Cf neutron source (right). The top panel shows that the X-Ray produce point like hits, with size determined by the charge diffusion inside the detector. The bottom panel shows the diffusion limited hits from nuclear recoils produced by the neutrons, together with Compton scattered electrons producing tracks.

Recent advances in CCD technology [12] allow the fabrication of high resistivity (~10kΩ/cm2 ) detectors, up to ~300 μm thick which are fully depleted at relatively low bias voltages. These CCDs have a significantly higher quantum efficiency in the near infra-red wavalengths than conventional, thin CCDs. For this reason these CCDs are the optical detectors chosen by several groups building new mosaic cameras for astronomy, such as DECam [13] , SNAP [14] and HyperSuprime [15]. For a general description of CCDs see Ref [16]. We consider here the CCDs developed [12] by Lawrence Berkeley National Laboratory for DECam [13] [17] [18]. These devices are 63 x 33 mm2 and have square pixels 15 microns on each side. The CCDs weigh 1 gram. The detectors are p-channel devices thinned to 250 microns thickness. Application of bias voltage on the back side of the CCD establishes a depletion region. Interactions in the depletion region produce electron-hole pairs. Positive-charge holes traverse the silicon and are collected in the buried channels a few microns from the gate electrodes.

The noise performance of these detectors has been studied in detail as part of the characterization effort done by the DECam CCD team [17] [18]. The detectors have 8.4 million pixels arranged in 2048 columns and 4096 rows, and are read out by two amplifiers in parallel. Each amplifier sitting on opposite ends of a serial register towards which the charge is clocked from the center of the row. The detectors have an output stage with a electronic gain of ~2.5 μV/e-. The signal is digitized after correlated double sampling (CDS) of the output. The CDS removes the noise resulting from resetting the output stage after each pixel, and is also efficient rejecting the common mode noise (frequencies much lower than the pixel frequency). Each sample used for the CDS operation is the result of an integration during a time τ. The integration time can be selected freely and acts as a filter for high frequency noise. The noise observed for pixel readout times larger than 50 μs is σ < 2 e- (RMS). These results were obtained using a Monsoon [19] CCD controller.

As charge is produced on the back of the CCD, it has to travel to the potential minimum a few microns away from the gate electrodes, during this time it can diffuse to neighbor pixels. When considering these detectors for a dark matter search, lateral charge diffusion is an important parameter because it determines the size of the reconstructed nuclear recoil events. The nuclear recoils will produce a very localized charge cloud ( << 15 μm), and the signature in the CCD detectors will be a diffusion-limited charge deposition. Photons in the few keV energy range (X-rays) will also produce a diffusion limited hit, and for this reason we use X-ray hits in a back illuminated CCD to measure diffusion. Diffusion measurements were done for DECam CCDs using X-rays and optical methods [20], diffusion was also studied on similar CCDs by  LBNL [21] [22]. The results indicate ~7 μm lateral charge diffusion.

Figure 2 shows an example of an X-ray image with diffusion limited hits (top) and an image obtained with a CCD exposure to neutrons (bottom) with diffusion limited hits, and other types of radiation. The size of the reconstructed hits on the CCD provides us with a tool for selection nuclear recoils in our images.

In most models, the interaction of a DM particle with a CCD detector will produce a nuclear recoil. For this reason is it very important to understand the ionization efficiency for nuclear recoils in these detectors. Some of the recoil energy is lost as acoustic vibrations or thermal excitations, the rest produces e-hole pairs. This so-called quenching factor must be accounted for when making measurements of nuclear recoil energy. The quenching factor has historically been well-described by the Lindhard theory [23]. The Lindhard theory has been applied in recent work [24] [25] and the quenching factor has been measured for silicon at energies above 4 keV [26]. However, at the low-energy region where low-mass DM candidates could be detected, the quenching becomes more dramatic and is increasingly energy-dependent. During the summer of 2009 we exposed a DECam CCD to a $^{252}$Cf neutron source to produce a preliminary measurement quenching factor at low energies, the results are shown in Figure 3.

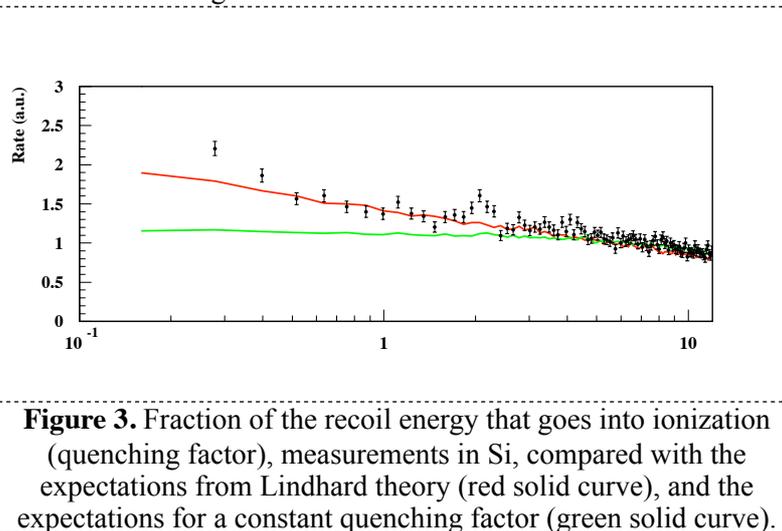

**Figure 3.** Fraction of the recoil energy that goes into ionization (quenching factor), measurements in Si, compared with the expectations from Lindhard theory (red solid curve), and the expectations for a constant quenching factor (green solid curve).

During the winter of 2009 we conducted a first test with DECam CCDs underground inside a lead shield. The idea for this test was to study the performance of the detectors in a low background environment. For this test we built an 8" lead shield using Fermilab surplus lead for the outer 6" and new lead, with low $^{210}$Pb content, in the inner 2". We operated the detectors inside a vacuum vessel that was available at Fermilab from previous work, this vessel was made out of stainless steel and its shape made shielding very inefficient. The detectors were operated at -140 C, and kept at that temperature using a cryo-cooler available from Fermilab surplus. The electronics used for this experiment were engineering versions of the DECam electronics, including engineering grade DECam CCDs. We build a clean tent in the Minos near detector hall for the installation of this apparatus.

For this run, we achieved two orders of magnitude reduction in background compared with the tests of the CCDs done at the surface in Lab-A at Fermilab. The measured rate of radiation events was $10^4$ cpd/kg/keV for nuclear recoils (diffusion limited hits) in the region 0.1-5 keV. This rate is still three orders of magnitude larger than what most low background experiments have achieved. This test was very useful to understand the background in our existing setup, and in this way establish a benchmark for the future development. We are also using the data obtained to understand the noise stability of the readout system, possible light leaks and the idea of rejecting non-DM events by looking at coincidences of hits in our 4 CCD array. For comparison with other experiments we present our results as a cross section limit as a function of dark matter particle mass. The limits, presented in Fig. 5, show that our 2009 run is about 2 orders of magnitude above the best existing limit for masses of 1 GeV. We also present in the figure the expectation for future DAMIC runs, assuming that we achieve 10 cpd/keV/kg and no improvements in the electronic readout noise (blue), and with a factor of 4 improvement in the electronic readout noise (yellow). These projections are calculated assuming 300 g-day exposure, which we could achieve with a 10 g detector ( 8 CCD array).

The tests performed underground during 2009 used a surplus vacuum vessel made out of stainless steel and used for earlier projects at Fermilab. The materials in this dewar were not selected considering their radio purity. In addition to this, the geometrical form of this vacuum vessel makes the shielding very inefficient, requiring a lot of shielding material and having to maintain inside the shield unnecessary components like the vacuum gauges and the cryogenic cooler head. We are now starting to build a low background dewar for this experiment and expect to start demonstrating the true potential of this technology. The recent results from DAMA/Libra have produced a generalized interest in the low mass DM searches. There results have been accompanied by the development of DM models which also point to low a mass dark matter particle. We are proposing here an experiment with CCDs to search for such low mass dark matter particle. Other groups are also conducting or planning dark matter searches for low mass particles [27][28]. The sensitivity of our experiments to a dark matter particle will depend on the results of the planned R&D tasks for lowering electronic readout noise and radiation background. We believe that this technique has great potential for low mass dark matter searches, and for other experiments requiring low threshold low background detectors.

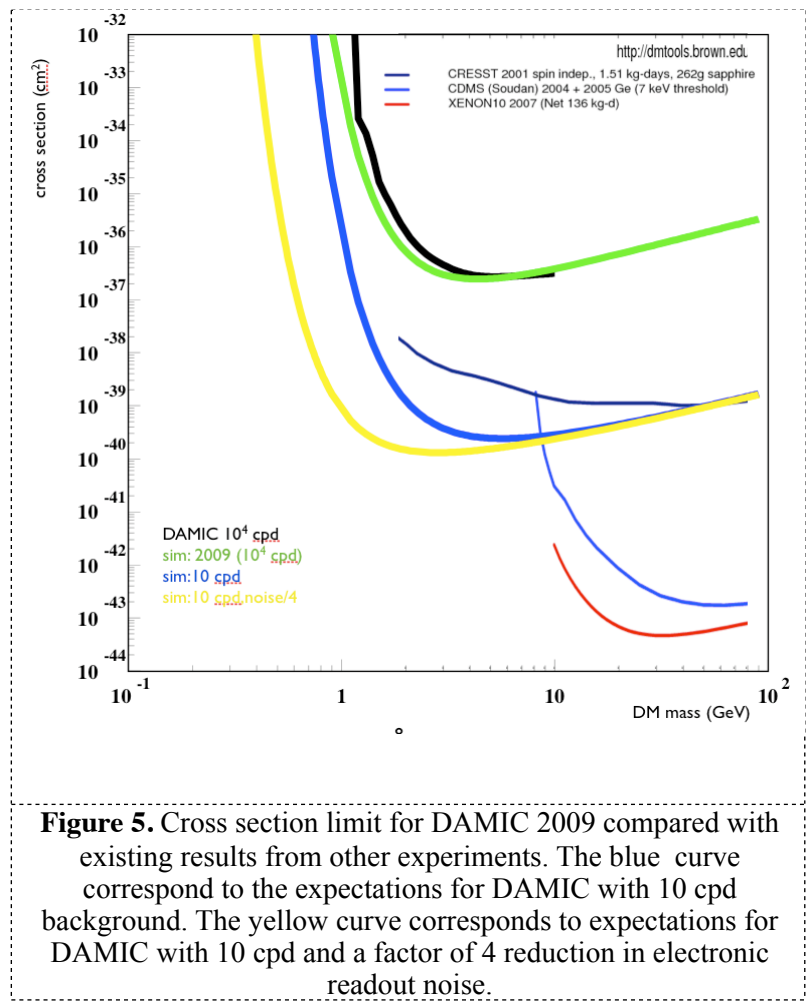

**Figure 5.** Cross section limit for DAMIC 2009 compared with existing results from other experiments. The blue curve correspond to the expectations for DAMIC with 10 cpd background. The yellow curve corresponds to expectations for DAMIC with 10 cpd and a factor of 4 reduction in electronic readout noise.

[1] R. J. Gaitskell, Annu. Rev. Nucl. Part. Sci., 54 315 (2004).
[2] G.Chardin, "Cryogenic Particle Detection", edited by Christian Enss, Springer (2005), arXiv:astro-ph/0411503
[3] P.S. Barbeau, J.I .Collar and O. Tench, Journal of Cosmology and Astroparticle Physics, 09 009 (2007).
[4] B. W. Lee and S. Weinberg, Phys. Rev. Lett. , 39 165 (1977).
[5] D. Hooper and K.M. Zurek, arXiv:0801.3686v1
[6] X. He, T. Li, X. Li and H. Tsai, Modern Physics Letters A, 22 2121 (2007).
[7] J.F. Gunion, D. Hooper, B. McElrath, Phys.Rev., D73 015011 (2006) .
[8] C. Bird, R. Kowalewski \& M. Pospelov, Modern Physics Letters A, 21 457 (2006).
[9] Gondolo, P. and Gelmini, G., Phys. Rev. , D71 123520 (2005).
[10] T. Damour and L.M. Krauss, ``Proceedings of the 3rd International Workshop on the Identification of Dark Matter", edited by N. J. C. Spooner & V. Kudryavtsev. World Scientific (2001), arXiv:astro-ph/9806165v3
[11] F. Petriello and K. M. Zurek, JHEP 0809, 047 (2008) [arXiv:0806.3989 [hep-ph]]; C. Savage, G. Gelmini, P. Gondolo and K. Freese, arXiv:0808.3607 [astro-ph].
C. Savage, K. Freese, P. Gondolo and D. Spolyar, arXiv:0901.2713 [astro-ph].
[12] S.E. Holland, D.E. Groom, N.P. Palaio, R. J. Stover, and M. Wei, IEEE Trans. Electron Dev., 50 225 (2003), LBNL-49992.
[13] Dark Energy Survey Collaboration, astro-ph/0510346; Flaugher, B., Ground-based and Airborne Instrumentation for Astronomy. Edited by McLean, Ian S.; Iye, Masanori. Proceedings of the SPIE, Volume 6269, (2006)
[14] ``Supernova / Acceleration Probe: A Satellite Experiment to Study the
  Nature of the Dark Energy", SNAP Collaboration, G.~Aldering et al.,
  submitted to Publ. Astr. Soc. Pac., astro-ph/0405232; SNAP Collaboration, astro-ph/0507459.
[15] M. Satoshi et al., Ground-based and Airborne Instrumentation for Astronomy. Edited by McLean, Ian S.; Iye, Masanori. Proceedings of the SPIE, Volume 6269, (2006)
[16] J.R. Janesick, Scientific Charge Cupled Devices, SPIE press (2001).
[17] J. Estrada & R. Schmidt , Scientific Detectors for Astronomy 2005, Edited by J.E. Beletic, J.W. Beletic and P. Amico, Springer, (2006).
[18] J. Estrada et al. , Ground-based and Airborne Instrumentation for Astronomy. Edited by McLean, Ian S.; Iye, Masanori. Proceedings of the SPIE, Volume 6269, (2006).
[19] http://www.noao.edu/ets/new_monsoon
[20] H. Cease, H. T. Diehl, J. Estrada, B. Flaugher and V. Scarpine, Experimental Astronomy, Online First (2007).
[21] J.A. Fairfield , D. E. Groom, S. J. Bailey, C. J. Bebek, S. E. Holland, A. Karcher, W. F. Kolbe, W. Lorenzon, & N. A. Roe, Fairfield IEEE Trans. Nucl. Sci. 53 (6), 3877 (2006)
[22] A. Karcher, C.J. Bebek, W. F. Kolbe, D. Maurath, V. Prasad, M. Uslenghi, M. Wagner, IEEE Trans. Nucl. Sci. 51 (5), (2004) LBNL-55685
[23] Lindhard, J et al, 1963 *Mat. Fys. Medd. Dan. Vid. Selsk.* 33.10
[24] Chagani, H et al, 2008 *JINST* 3 P06003
[25] Lewin J & Smith, P, 1996 *P Astropart. Phys.* 6 87.
[26] Gerbier, G. et al., Phys.Rev.D. 42, 3211 (1990).
[27] S.T. Lin et al (Texono Collaboration), Phys.Rev.D79:061101,2009.
[28] Aalseth, C.E. et al, Review Letters, vol. 101, Issue 25, id. 251301